\documentclass[preprint,preprintnumbers,10pt,byrevtex,aps,prl,twocolumn,tightenlines,superscriptaddress,
showpacs,amsmath,amssymb]{revtex4}

\usepackage{graphicx} 
\usepackage{dcolumn} 
\usepackage{color}
\usepackage{epstopdf}

\usepackage{xspace}
\newcommand{\acp}{\ensuremath{\mathcal{A}_{CP}}\xspace}
\newcommand{\BCG}{\ensuremath{\bar{B}\rightarrow X_{c}\gamma}\xspace}
\newcommand{\BUG}{\ensuremath{\bar{B}\rightarrow X_{u}\gamma}\xspace}
\newcommand{\BSG}{\ensuremath{\bar{B}\rightarrow X_{s}\gamma}\xspace}
\newcommand{\BDG}{\ensuremath{\bar{B}\rightarrow X_{d}\gamma}\xspace}
\newcommand{\BSDG}{\ensuremath{\bar{B}\rightarrow X_{s+d}\gamma}\xspace}
\newcommand{\BbSDG}{\ensuremath{B\rightarrow X_{\bar{s}+\bar{d}}\gamma}\xspace}
\newcommand{\xsd}{\ensuremath{X_{s+d}}\xspace}
\newcommand{\bsg}{\ensuremath{b\rightarrow s\gamma}\xspace}
\newcommand{\bdg}{\ensuremath{b\rightarrow d\gamma}\xspace}
\newcommand{\BBbar}{\ensuremath{B\bar{B}}\xspace}
\newcommand{\epem}{\ensuremath{e^+e^-}\xspace}

\newcommand{\ecmg}{\ensuremath{E^{\rm{*}}_{\gamma}}\xspace}
\newcommand{\pcml}{\ensuremath{p^{\rm{*}}_{\ell}}\xspace}
\newcommand{\ups}{\ensuremath{\Upsilon(4S)}\xspace}
\newcommand{\alid}{\ensuremath{\mathcal{A}_{\rm{LID}}}\xspace}
\newcommand{\adet}{\ensuremath{\mathcal{A}_{\rm{det}}}\xspace}
\newcommand{\abkg}{\ensuremath{\mathcal{A}_{\rm{bkg}}}\xspace}
\newcommand{\atrack}{\ensuremath{\mathcal{A}_{\rm{track}}}\xspace}
\newcommand{\GeV}{\ensuremath{\rm{GeV}}\xspace}

\begin{document}

\preprint{\vbox{ \hbox{ }
 \hbox{Belle preprint {\it 2014-21 }}
 \hbox{KEK preprint {\it 2014-37 }}
}}

\title{ \quad\\[1.0cm] Measurement of the direct $CP$ asymmetry in \BSDG\\decays with a lepton tag}

\begin{abstract}
We report the measurement of the direct $CP$ asymmetry in the radiative \BSDG decay using a data sample of $(772 \pm 11)\times 10^6$ \BBbar pairs collected at the \ups resonance with the Belle detector at the KEKB asymmetric-energy \epem collider. The $CP$ asymmetry is measured as a function of the photon energy threshold. For $\ecmg \geq 2.1~{\GeV}$, where $\ecmg$ is the photon energy in the center-of-mass frame, we obtain $\acp(\BSDG)= (2.2 \pm 3.9 \pm 0.9) \%$, consistent with the Standard Model prediction.
\end{abstract}

\noaffiliation
\affiliation{University of the Basque Country UPV/EHU, 48080 Bilbao}
\affiliation{Beihang University, Beijing 100191}
\affiliation{University of Bonn, 53115 Bonn}
\affiliation{Budker Institute of Nuclear Physics SB RAS and Novosibirsk State University, Novosibirsk 630090}
\affiliation{Faculty of Mathematics and Physics, Charles University, 121 16 Prague}
\affiliation{University of Cincinnati, Cincinnati, Ohio 45221}
\affiliation{Deutsches Elektronen--Synchrotron, 22607 Hamburg}
\affiliation{Justus-Liebig-Universit\"at Gie\ss{}en, 35392 Gie\ss{}en}
\affiliation{The Graduate University for Advanced Studies, Hayama 240-0193}
\affiliation{Hanyang University, Seoul 133-791}
\affiliation{University of Hawaii, Honolulu, Hawaii 96822}
\affiliation{High Energy Accelerator Research Organization (KEK), Tsukuba 305-0801}
\affiliation{IKERBASQUE, Basque Foundation for Science, 48013 Bilbao}
\affiliation{Indian Institute of Technology Bhubaneswar, Satya Nagar 751007}
\affiliation{Indian Institute of Technology Guwahati, Assam 781039}
\affiliation{Indian Institute of Technology Madras, Chennai 600036}
\affiliation{Institute of High Energy Physics, Chinese Academy of Sciences, Beijing 100049}
\affiliation{Institute of High Energy Physics, Vienna 1050}
\affiliation{Institute for High Energy Physics, Protvino 142281}
\affiliation{INFN - Sezione di Torino, 10125 Torino}
\affiliation{Institute for Theoretical and Experimental Physics, Moscow 117218}
\affiliation{J. Stefan Institute, 1000 Ljubljana}
\affiliation{Kanagawa University, Yokohama 221-8686}
\affiliation{Institut f\"ur Experimentelle Kernphysik, Karlsruher Institut f\"ur Technologie, 76131 Karlsruhe}
\affiliation{Kennesaw State University, Kennesaw GA 30144}
\affiliation{Department of Physics, Faculty of Science, King Abdulaziz University, Jeddah 21589}
\affiliation{Korea Institute of Science and Technology Information, Daejeon 305-806}
\affiliation{Korea University, Seoul 136-713}
\affiliation{Kyungpook National University, Daegu 702-701}
\affiliation{\'Ecole Polytechnique F\'ed\'erale de Lausanne (EPFL), Lausanne 1015}
\affiliation{Faculty of Mathematics and Physics, University of Ljubljana, 1000 Ljubljana}
\affiliation{Luther College, Decorah, Iowa 52101}
\affiliation{University of Maribor, 2000 Maribor}
\affiliation{Max-Planck-Institut f\"ur Physik, 80805 M\"unchen}
\affiliation{School of Physics, University of Melbourne, Victoria 3010}
\affiliation{Moscow Physical Engineering Institute, Moscow 115409}
\affiliation{Moscow Institute of Physics and Technology, Moscow Region 141700}
\affiliation{Graduate School of Science, Nagoya University, Nagoya 464-8602}
\affiliation{Kobayashi-Maskawa Institute, Nagoya University, Nagoya 464-8602}
\affiliation{Nara Women's University, Nara 630-8506}
\affiliation{National Central University, Chung-li 32054}
\affiliation{National United University, Miao Li 36003}
\affiliation{Department of Physics, National Taiwan University, Taipei 10617}
\affiliation{H. Niewodniczanski Institute of Nuclear Physics, Krakow 31-342}
\affiliation{Niigata University, Niigata 950-2181}
\affiliation{Osaka City University, Osaka 558-8585}
\affiliation{Pacific Northwest National Laboratory, Richland, Washington 99352}
\affiliation{Peking University, Beijing 100871}
\affiliation{University of Pittsburgh, Pittsburgh, Pennsylvania 15260}
\affiliation{University of Science and Technology of China, Hefei 230026}
\affiliation{Seoul National University, Seoul 151-742}
\affiliation{Soongsil University, Seoul 156-743}
\affiliation{Sungkyunkwan University, Suwon 440-746}
\affiliation{School of Physics, University of Sydney, NSW 2006}
\affiliation{Department of Physics, Faculty of Science, University of Tabuk, Tabuk 71451}
\affiliation{Tata Institute of Fundamental Research, Mumbai 400005}
\affiliation{Excellence Cluster Universe, Technische Universit\"at M\"unchen, 85748 Garching}
\affiliation{Toho University, Funabashi 274-8510}
\affiliation{Tohoku University, Sendai 980-8578}
\affiliation{Department of Physics, University of Tokyo, Tokyo 113-0033}
\affiliation{Tokyo Institute of Technology, Tokyo 152-8550}
\affiliation{Tokyo Metropolitan University, Tokyo 192-0397}
\affiliation{University of Torino, 10124 Torino}
\affiliation{CNP, Virginia Polytechnic Institute and State University, Blacksburg, Virginia 24061}
\affiliation{Wayne State University, Detroit, Michigan 48202}
\affiliation{Yamagata University, Yamagata 990-8560}
\affiliation{Yonsei University, Seoul 120-749}

  \author{L.~Pes\'{a}ntez}\affiliation{University of Bonn, 53115 Bonn} 
  \author{P.~Urquijo}\affiliation{School of Physics, University of Melbourne, Victoria 3010} 
  \author{J.~Dingfelder}\affiliation{University of Bonn, 53115 Bonn} 
  
  \author{A.~Abdesselam}\affiliation{Department of Physics, Faculty of Science, University of Tabuk, Tabuk 71451} 
  \author{I.~Adachi}\affiliation{High Energy Accelerator Research Organization (KEK), Tsukuba 305-0801}\affiliation{The Graduate University for Advanced Studies, Hayama 240-0193} 
  \author{K.~Adamczyk}\affiliation{H. Niewodniczanski Institute of Nuclear Physics, Krakow 31-342} 
  \author{H.~Aihara}\affiliation{Department of Physics, University of Tokyo, Tokyo 113-0033} 
  \author{S.~Al~Said}\affiliation{Department of Physics, Faculty of Science, University of Tabuk, Tabuk 71451}\affiliation{Department of Physics, Faculty of Science, King Abdulaziz University, Jeddah 21589} 
  \author{K.~Arinstein}\affiliation{Budker Institute of Nuclear Physics SB RAS and Novosibirsk State University, Novosibirsk 630090} 
  \author{D.~M.~Asner}\affiliation{Pacific Northwest National Laboratory, Richland, Washington 99352} 
% ...
\collaboration{The Belle Collaboration}

\date{\today}

\pacs{11.30.Er, 13.25.Hw}

\maketitle
	
{\renewcommand{\thefootnote}{\fnsymbol{footnote}}}
\setcounter{footnote}{0}

The radiative electroweak transitions \bsg and \bdg proceed via flavor-changing neutral currents involving loop diagrams. These decays are sensitive to possible contributions from new heavy particles occurring in the loop, which modify the branching fractions and $CP$-violating effects predicted in the Standard Model (SM). The decay rates, including QCD corrections, can be expressed by an effective Hamiltonian and calculated using the Operator Product Expansion approach. In the leading and next-to-leading order logarithmic approximation, the branching fractions and $CP$ asymmetries are proportional to the dipole operators $P_{7}$ and $P_{8}$~\cite{Gambino:2001ew}. New physics effects would modify the corresponding Wilson coefficients $C_{7}$ and $C_{8}$.

The $CP$ asymmetry (\acp) in \BSDG decays is defined as:
\begin{equation}
\acp(\BSDG) \equiv \frac{\Gamma(\BSDG)-\Gamma(\BbSDG)}{\Gamma(\BSDG)+\Gamma(\BbSDG)},
\label{eq:acp_width}
\end{equation}
where $\Gamma(\BSDG)$ represents the decay rate of the $\overline{B^{0}}$ or $B^{-}$ meson into the radiative final state. In the following, charge-conjugate states are included implicitly. The \xsd states represent all possible hadronic final states derived from \bsg or \bdg transitions. The SM predicts \acp for the these two transitions in the ranges $-0.6\%\leq\acp(\BSG)\leq 2.8\%$ and $-62\%\leq \acp(\BDG)\leq 14\%$~\cite{Benzke:2010tq}. Even though the individual $CP$-violating effects could be large, the $CP$-violating contributions cancel  when both are considered inclusively due to CKM unitarity, and the theory errors cancel almost perfectly except for small $U$-spin breaking corrections~\cite{Hurth:2003dk}, additionally, the inclusive asymmetry is insensitive to the choice of photon energy cutoff~\cite{Kagan:1998bh}. This precise SM prediction of \acp(\BSDG)=0, serves as a clean test for new $CP$-violating phases acting in the decays. New physics (NP) scenarios such as supersymmetric models with minimal flavor violation predict \acp(\BSDG) up to a level of $+2~\%$. In more generic NP scenarios, the asymmetries \acp(\BSG) and \acp(\BDG) do not cancel and \acp(\BSDG) is the most sensitive observable, with values as large as $10\%$~\cite{Hurth:2003dk}.

Previous measurements of \acp(\BSDG) have been performed by CLEO~\cite{Coan:2000pu} and BaBar~\cite{Lees:2012ufa} and are statistically limited. Belle has performed a measurement of the inclusive branching fraction~\cite{Limosani:2009qg}. The asymmetry \acp(\BSG) has been measured separately as the sum of exclusive decays~\cite{Nishida:2003paa,Lees:2014uoa}. In this letter, we present the first Belle measurement of $\acp(\BSDG)$. We profit from the large data sample available at Belle to achieve a higher statistical precision.

The states \xsd include resonant contributions such as $K^{*}(892)$, $\rho$ and $\omega$, and non-resonant contributions. In order to be sensitive to all \xsd states, the selection is based on the high-energy-photon signature of the transition, \textit{i.e.} the radiated photon is the only reconstructed particle from the \BSDG decay. While this approach does not exclude explicitly possible contributions from \BCG or \BUG decays, such contributions are very small in the SM~\cite{Cheng:1994qw} and will be neglected in this analysis. To tag the signal $B$ flavor, we use the fact that $B$ mesons are produced in pairs from the reaction $\epem \rightarrow \ups \rightarrow \BBbar$. The flavor of the signal $B$ meson is determined by tagging the flavor of the other $B$ in the event, using a charged lepton ($e,\,\mu$) consistent with the semileptonic decay of the other $B$. The $B$ flavor and lepton charge in semileptonic decays are directly related.

Since the expected $CP$ violation is very small and precisely calculable, all effects that could bias the measurement must be carefully quantified. A measurement bias is introduced if the selection procedure, track reconstruction, or particle identification favors a particular charge. These effects are quantified in different control samples. In this analysis, we also test the independence of \acp with respect to the choice of cutoff energy, by measuring it as a function of the photon energy threshold.

This analysis uses the $711\rm{~fb}^{-1}$ sample recorded at the \ups resonance by the Belle experiment at the KEKB storage ring~\cite{KEKB}, containing $(772 \pm 11)\times 10^6$ \BBbar pairs. An $89\rm{~fb}^{-1}$ sample recorded at a center-of-mass (CM) energy $60$~MeV below the resonance is used to study continuum background ($\epem \rightarrow q\bar{q}$, where $q= u,d,s,c$); the former sample is denoted on-resonance and the latter off-resonance. The Belle detector is a large-solid-angle magnetic spectrometer that consists of a silicon vertex detector (SVD), a 50-layer central drift chamber (CDC), an array of aerogel threshold Cerenkov counters (ACC), a barrel-like arrangement of time-of-flight scintillation counters (TOF), and an electromagnetic calorimeter comprised of CsI(Tl) crystals (ECL) located inside a super-conducting solenoid coil that provides a $1.5$~T magnetic field. An iron flux-return located outside of the coil is instrumented to detect $K_L^0$ mesons and to identify muons (KLM). The detector is described in detail elsewhere~\cite{Abashian:2000cg}.

Monte Carlo (MC) simulation samples were generated to study continuum background, \BBbar decays and \BSG signal events. The size of the \BBbar MC sample is equivalent to ten times the integrated luminosity of the data. The size of the continuum MC sample corresponds to the integrated luminosity of the on-resonance sample. The generation of the signal \BSG decays follows the theoretical prediction of the Kagan-Neubert model~\cite{Kagan:1998ym} with parameters $m_{b} = 4.574~{\GeV}$ and $\mu_{\pi}^2 = 0.459~{\GeV}^2$ representing the $b$-quark mass and mean kinetic energy. The signal sample contains 2.6 million \BSG events, which corresponds roughly to five times the number expected in data. The \BBbar and \BSG MC samples included $B^0$-$\bar{B}^0$ mixing.

In this analysis, tracks passing very far from the interaction point or compatible with a low-momentum particle reconstructed multiple times as it spirals in the CDC are excluded. For photons, minimum energies of $100~\rm{MeV}$, $150~\rm{MeV}$ and $50~\rm{MeV}$,respectively, are required in the forward, backward and barrel regions of ECL, defined in Ref.~\cite{Abashian:2000cg}. These requirements suppress low-energy photons resulting from particle interactions with detector material or the beam pipe. All particles are used to calculate kinematic and topological variables.

The signal photon candidates are selected as connected clusters of ECL crystals in the polar angle $32.2^{\circ}\leq \theta_{\gamma}\leq 128.7^{\circ}$ with a CM energy $1.4~{\GeV}\leq \ecmg \leq 4.0~{\GeV}$. The polar angle is measured from the $z$ axis that is collinear with the positron beam. The ratio of the energy deposit in the central $3\times3$ crystals to that in the central $5\times5$ crystals must be larger than $90\%$. Photons from the decays $\pi^0(\eta)\rightarrow \gamma\gamma$ are rejected by using a veto based on the photon energy, polar angle and the reconstructed diphoton mass, as described in Ref.~\cite{Koppenburg:2004fz}. The signal region includes photons with CM energy $1.7~{\GeV}\leq \ecmg \leq 2.8~{\GeV}$; the sidebands $\ecmg < 1.7~{\GeV}$ and $\ecmg > 2.8~{\GeV}$ are used to study the normalization of \BBbar and continuum background components, respectively.

The lepton candidates used for tagging (tag lepton) are reconstructed as tracks in the SVD and CDC. We limit the impact parameters along the $z$ axis to $|dz| \leq 2~\rm{cm}$ and $dr \leq 0.5~\rm{cm}$, require at least one hit in the SVD, and choose a momentum range in the CM frame of $1.10~{\GeV}\leq \pcml\leq 2.25~{\GeV}$. The upper-momentum bound reduces continuum background as it is near the kinematic limit for leptons from $B$ decays. The lower bound ensures that most of the selected leptons originate directly from a $B$ meson, which is important for flavor tagging. Electron candidates are identified by constructing a likelihood ratio based on the matching of the cluster in the ECL and the extrapolated track, the ratio between its energy and momentum, the shower shape in the ECL, the energy loss in the CDC, and the light yield in the ACC. The polar angle requirement for electrons is $18^{\circ}\leq \theta_{e}\leq 150^{\circ}$. Muon identification uses a likelihood ratio determined from the range of the track and the normalized transverse deviations between the track and the KLM hits associated to it. The polar angle requirement for muons is $25^{\circ}\leq \theta_{\mu}\leq 145^{\circ}$.

\begin{figure*}[htb]
\includegraphics[width=0.45\textwidth]{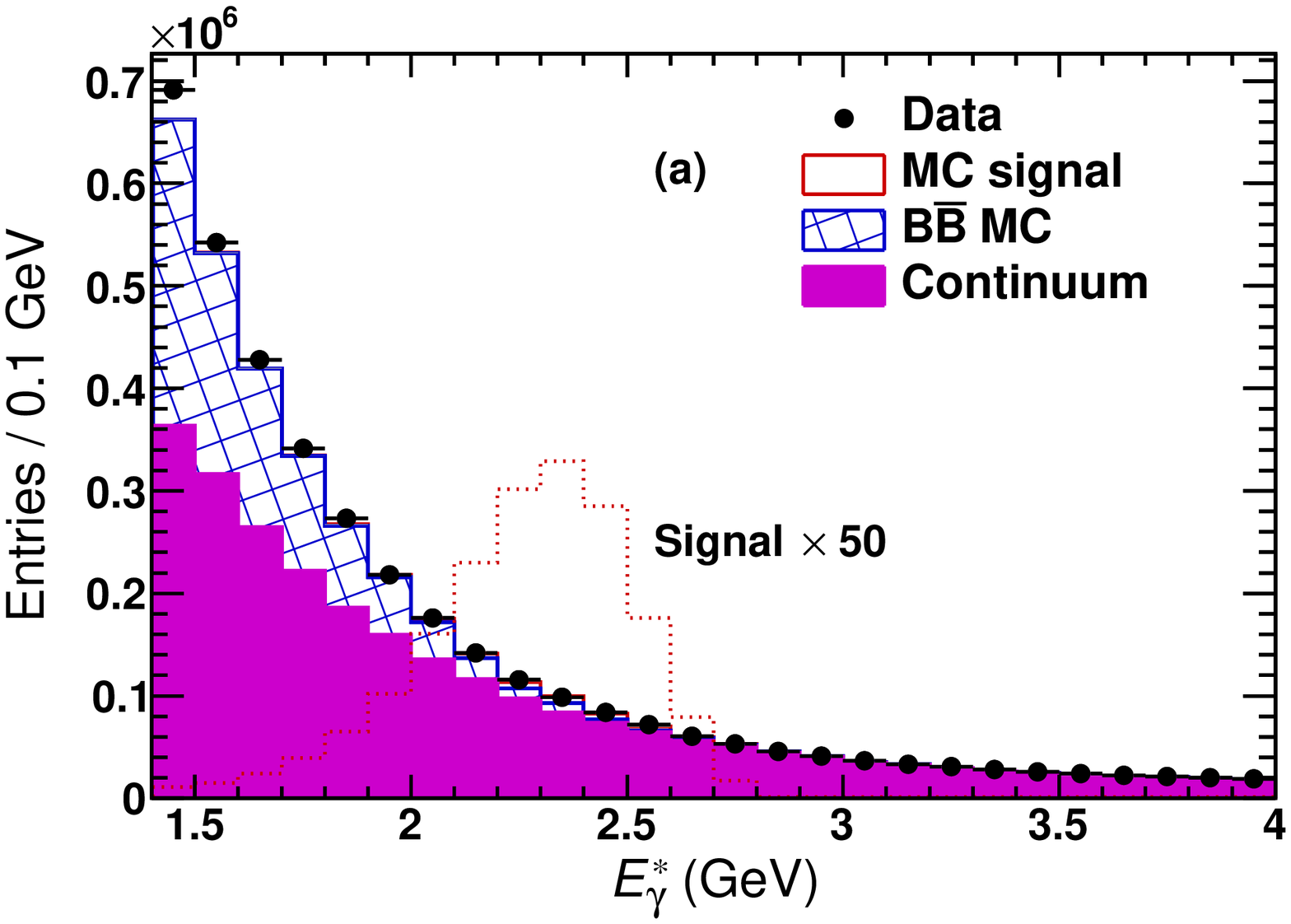}
\includegraphics[width=0.45\textwidth]{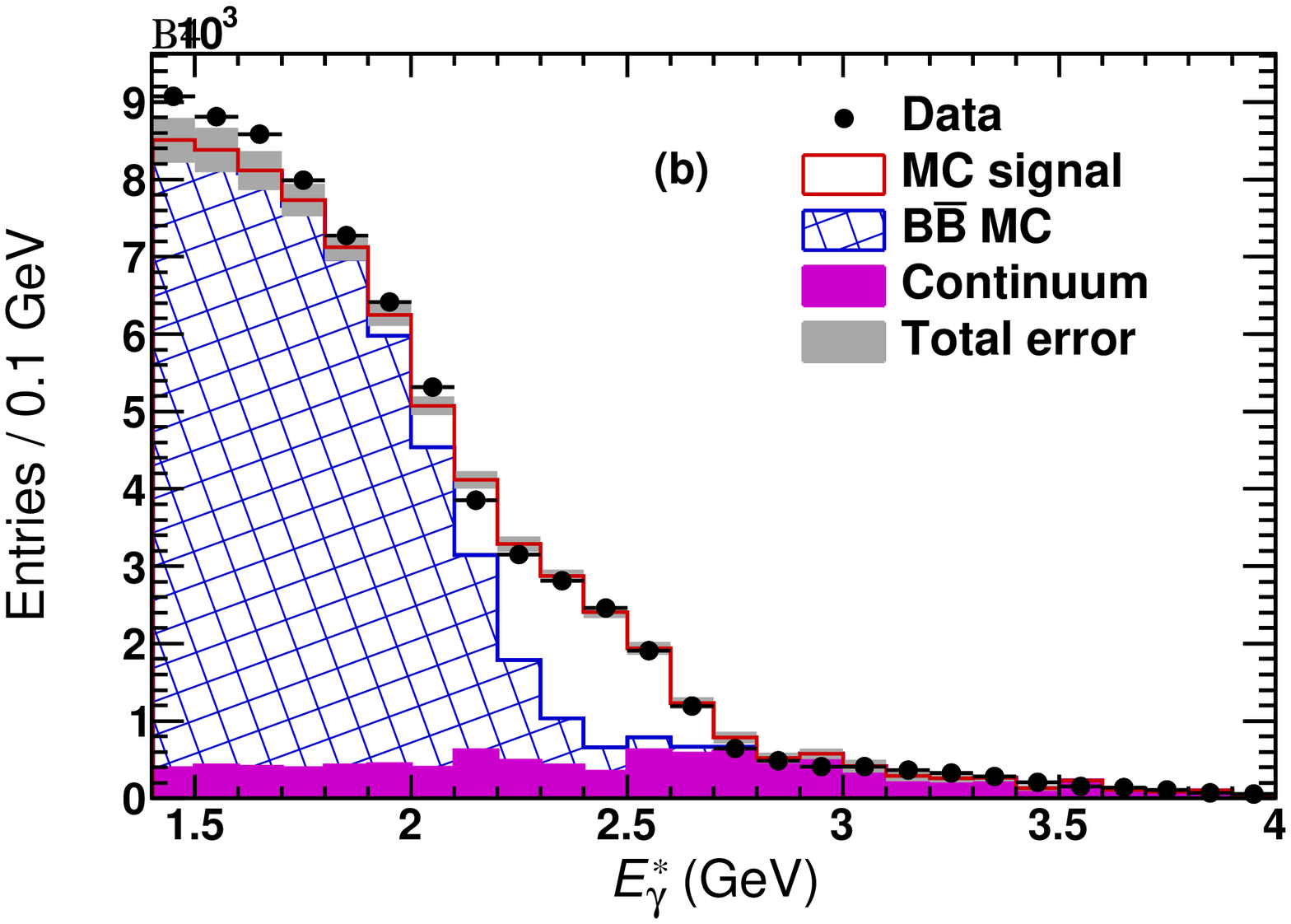}
\caption{Photon energy spectrum in the CM frame showing on-resonance data, off-resonance data for continuum, and MC simulation. The spectrum is shown (a) before and (b) after continuum suppression. In (a), the MC signal is additionally plotted scaled a factor of fifty to show its expected position. In (b), the MC error includes statistical and systematic uncertainties coming from calibration and normalization factors that cancel in the measurement of \acp.}
\label{fig:spectrum}
\end{figure*}

\begin{figure}[htb]
\includegraphics[width=0.45\textwidth]{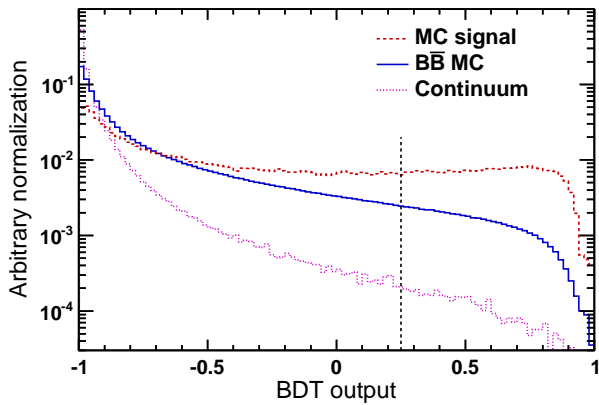}
\caption{Output of the BDT. The continuum distribution corresponds to off-resonance data. The vertical line denotes the minimum requirement on this variable.}
\label{fig:bdt}
\end{figure}

After this initial selection, the sample is dominated by continuum background, which constitutes $77\%$ of the total yield; the signal component amounts only to $1\%$ as can be seen in Fig.~\ref{fig:spectrum}(a). To suppress the continuum background, we use a Boosted Decision Tree (BDT), that is trained to achieve the best discrimination between continuum and signal events. Eighteen kinematic, event shape and isolation variables are used as input for the BDT: eleven Fox-Wolfram moments~\cite{Fox:1978vu}, constructed in three sets in which (1) all particles in the event are used, (2) the signal photon is excluded and (3) both signal photon and tag lepton are excluded; the magnitude and direction of the event's thrust vector; the distance between the photon cluster and the closest extrapolated position of a charged particle at the ECL surface; the angle between the directions of the photon and tag lepton; the RMS width of the photon cluster; the scalar sum of the transverse momenta of all reconstructed particles; and the square of the missing four-momentum, calculated as the difference between the total beam energy and the momenta of all reconstructed particles. The BDT is trained using continuum and \BSG MC samples. The selection criterion on the BDT output classifier variable is chosen to minimize the expected statistical uncertainty on \acp. The BDT classifier distribution and selection criterion are shown in Fig.~\ref{fig:bdt}. The photon spectrum after continuum suppression is shown in Fig.~\ref{fig:spectrum}(b) for MC and on-resonance data, in this plot we include statistical uncertainties and systematic uncertainties that come from calibration and normalization factors, that cancel in the measurement of \acp.

After the selection, in the region $1.7~{\GeV} \leq \ecmg \leq 2.8~{\GeV}$, we find 21400 (21608) events tagged with a positive (negative) lepton in the on-resonance sample and $2623 \pm 140$ ($2728 \pm 143$) events tagged with a positive (negative) lepton in the off-resonance sample. The off-resonance events are corrected as they have, on average, lower particle energies and multiplicities due to the lower CM energy. Additionally, the off-resonance yield is scaled to take into account the difference in luminosities and cross-sections.

The signal fraction is $21.2\%$ while the continuum background fraction is $12.4\%$. The \BBbar background contains photons from several processes. The dominant sources are photons from $\pi^0\rightarrow \gamma\gamma$ decays, which make up $49.5\%$ of the total yield and photons from $\eta\rightarrow \gamma\gamma$, contributing $7.9\%$. Photons from beam background are $2.2\%$ of the total contribution. Electrons and hadrons misidentified as photons are small contributions of $0.8\%$ and $0.2\%$, respectively. Other photons, mainly from decays of $\omega$, $\eta^{\prime}$ and $J/\psi$ mesons, and bremsstrahlung, including final state radiation~\cite{Barberio:1990ms}, comprise the remaining $5.8\%$. The \BSDG signal is obtained by subtracting the continuum and \BBbar contributions. The \BBbar background components are calibrated using data, as described below. All corrections and calibrations applied to MC and off-resonance data are determined and performed independently of the tag charge. The subtraction of background is done for each charge individually.

The rejection of events containing $\pi^{0}$ or $\eta$ will fail in cases where the decay is very asymmetric and the second photon has an energy below the threshold, making the reconstruction of the $\pi^{0}$ or $\eta$ impossible. To properly normalize these components, the veto is removed and, for each combination of the prompt photon with another photon in the event, the diphoton mass $m_{\gamma\gamma}$ is calculated. A fit to the $\pi^0$ and $\eta$ masses is performed to estimate the number of these mesons in data and MC. The fit is performed in eleven meson momentum bins between 1.4 and 2.6~\GeV and the ratio of data to MC yields is used as a correction factor.

Some background components have a non-vanishing direct $CP$ asymmetry that could impact our measurement. Most have negligible contributions to the decay rate except for $B \rightarrow X_{s}\eta$ decays, which comprises $1.2\%$ of the rate according to the MC prediction, with a branching fraction $\mathcal{B}(B \rightarrow X_{s}\eta) = \left(26.1\pm 3.0 \,^{+1.9} _{-2.1} \,^{+4.0} _{-7.1}\right) \times 10^{-5}$ and a $CP$ asymmetry $\acp(B \rightarrow X_{s}\eta)=(-13 \pm 5)\%$ measured by Belle~\cite{Nishimura:2009ae}. The MC is corrected to model this effect properly.

\begin{figure}[hhhtb]
\includegraphics[width=0.45\textwidth]{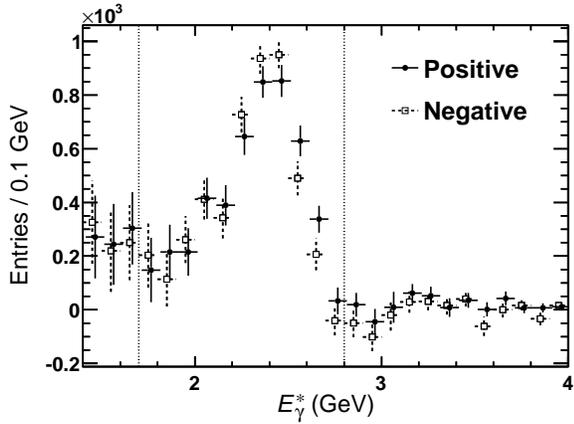}
\caption{The photon energy spectrum in the CM system after subtracting all the background, with vertical dashed lines showing the signal region. The positive tagged events are shown as circles and the negative as squares. Statistical and systematic uncertainties are included.}
\label{fig:subtracted}
\end{figure}

The \BSDG photon energy spectrum for positive and negative tagged events after subtracting all the background is shown in Fig.~\ref{fig:subtracted}. The measured asymmetry, $\acp^{\rm{meas}}$, is calculated using Eq.~(\ref{eq:acp_width}) expressed in terms of the charge-flavor correlation: $\acp^{\rm{meas}} = \dfrac{N^{+}-N^{-}}{N^{+}+N^{-}}$. Here, $N^{+}$ and $N^{-}$ represent the total number of events tagged by a positive or negative lepton for a given photon energy threshold. The energy thresholds range from $1.7$ to $2.2~{\GeV}$.

The measured values must be corrected due to possible asymmetries in the \BBbar background that is subtracted (\abkg) and possible asymmetries in the detection of leptons \adet. An additional correction arises from the probability that the reconstructed lepton has a wrong charge-flavor correlation, the so-called wrong-tag probability ($\omega$). The corrected asymmetry is given by:
\begin{equation}
\acp = \dfrac{1}{1-2\omega}( \acp^{\rm{meas}} - \abkg - \adet).
\label{eq:corrected}
\end{equation}
The correction \adet accounts for a possible asymmetry in the identification efficiency between positive and negative charged leptons (\alid) and a possible asymmetry between the reconstruction of positive and negative tracks (\atrack). \alid is determined using a $B \rightarrow X J/\psi( \ell^{+}\ell^{-})$ sample, where the selection efficiencies of positively and negatively charged electrons and muons are determined by performing fits to the invariant dilepton mass $m_{\ell\ell}$ for singly- and doubly-identified lepton candidates. The asymmetry is calculated as: $\alid=\dfrac{\varepsilon^{+}-\varepsilon^{-}}{\varepsilon^{+}+\varepsilon^{-}}$. This measurement is performed in the full kinematic region, in eleven laboratory-frame momentum bins and eight polar-angle bins. The asymmetries for electrons and muons are measured to be $\alid(e)=(0.26\pm 0.14)\%$ and $\alid(\mu)=(-0.03\pm 0.03)\%$, and average to $\alid=(0.11\pm 0.07)\%$. The asymmetry \atrack is measured with partially and fully reconstructed $D^{*}$ with $D^{*}\rightarrow \pi D^{0}$, $D^{0}\rightarrow \pi\pi K^{0}_{S}$, $K^{0}_{S}\rightarrow \pi^{+}\pi^{-}$ decays, to be $\atrack=(-0.01 \pm 0.21)\%$. The total detector-related asymmetry is $\adet=(0.10 \pm 0.22)\%$.

We measure \abkg in the low-energy sideband $\ecmg\leq 1.7~{\GeV}$. The asymmetries measured in data and MC are $\abkg(\rm{data})=(-0.14 \pm 0.78)\%$ and $\abkg(\rm{MC})=(-0.26 \pm 0.21)\%$, which are consistent with zero within uncertainties. The asymmetry in the \BBbar data is taken as a correction to \acp. Since this is an asymmetry in the \BBbar background, the correction is proportional to the ratio of \BBbar to signal events in the signal region, the ratios are taken from MC simulation.

The wrong-tag probability has contributions from $B^{0}\bar{B}^{0}$ oscillations ($\omega_{\rm{osc}}$), secondary leptons ($\omega_{\rm{sec}}$) and misidentified hadrons ($\omega_{\rm{misID}}$) and is given by $\omega=\omega_{\rm{osc}} + \omega_{\rm{sec}} + \omega_{\rm{misID}}$. The oscillation term is equal to the product of the mixing probability in the $B^{0}\bar{B}^{0}$ system $\chi_{d}=0.1875 \pm 0.0020$~\cite{Beringer:1900zz}, the fraction of neutral $B$ mesons from the $\Upsilon(4S)$ decay, $f_{00}=0.487 \pm 0.006$~\cite{Beringer:1900zz}, and the fraction of leptons coming directly from a $B$ decay, which is estimated to be $91.1\%$ from MC, resulting in $\omega_{\rm{osc}}=0.0832\pm 0.0015$. Secondary leptons are true leptons that do not come directly from a $B$ meson but rather from one of its decay daughters. We find $\omega_{\rm{sec}}=0.0431 \pm 0.0036$; this value is estimated from MC and the error based on the precision with which the $B\rightarrow D X$ and $D\rightarrow X l\nu$ branching fractions are measured. Misidentified hadrons give the smallest contribution and consist of $\pi$ and $K$ mesons faking a muon and, to a lesser extent, an electron. The corresponding wrong-tag probability is estimated from MC, where the fraction of misidentified hadrons is determined by studying $D^{*+}\rightarrow D^{0}(K^{-}\pi^{+}) \pi^{+}$ decays. After applying the same selection criteria for $\pi$ and $K$ candidates as for tag leptons, the fraction of hadrons passing the selection in the MC is corrected and we obtain $\omega_{\rm{misID}}=0.0069 \pm 0.0034$. The total wrong-tag probability value is $\omega = 0.1332 \pm 0.0052$.

The asymmetries \adet and \abkg are the dominant uncertainties on \acp and are additive. An additional multiplicative systematic uncertainty arises from the wrong-tag probability, leading to a relative uncertainty $\Delta \acp/\acp = 0.01$, much less than the additive uncertainties.

Finally, as some background events remain in the low energy range after subtraction, we scale the \BBbar component to match the data yield below $1.7$ GeV and recalculate \acp. The difference between this value and the nominal is taken as an additional systematic uncertainty.
	
In Table~\ref{table:acp:values}, the measured and corrected values of \acp are summarized for $0.1~{\GeV}$ steps in the \ecmg threshold from $1.7$ to $2.2~{\GeV}$, with a \ecmg upper bound of $2.8~{\GeV}$. The statistical and systematic uncertainties are summarized in Table~\ref{table:uncertainties}. The statistical precision is improved in comparison to previous measurements~\cite{Coan:2000pu,Lees:2012ufa}; it is, however, the limiting factor in the measurement and is affected by the size of the continuum sample. As an example, for the $1.7~{\GeV}$ threshold, the total $4.4\%$ statistical uncertainty incorporates a $3.0\%$ contribution from \ups data and $3.1\%$ from off-resonance data. The dominant systematic uncertainty arises from the asymmetry in the \BBbar background. The asymmetry is consistent with zero across the different photon energy thresholds.

\begin{table*}[hptb]
	\caption[]{$CP$ asymmetry, in percent, for different photon energy thresholds, the \ecmg kinematic limit is $2.8~{\GeV}$. For the measured asymmetry, only the statistical uncertainty is shown; for the corrected asymmetry \acp, statistical and systematic uncertainties are given. The ratio B/S representes the ratio in the number of \BBbar to signal events that is used to scale the asymmetry \abkg. An additional systematic uncertainty related to the wrong-tag probability is not explicitly listed but is taken into account in the total uncertainty; its relative value is $1\%$. The systematic contributions are added in quadrature.}
\centering
\begin{tabular}{lccccccc}
	\hline
$\ecmg(\rm{thresh.})$ & $\acp^{\rm{meas}}$ & B/S & \abkg & \adet & MC stats. & \BBbar norm. & \acp\\
 \hline 
$1.7~\rm{GeV}$ & $1.3 \pm 3.1$ & $3.20$ & $-0.4 \pm 2.5$ & $0.1 \pm 0.2$ & $\pm 0.8$ & $\pm 0.5$ & $2.2 \pm 4.3 \pm 3.5$ \\
$1.8~\rm{GeV}$ & $2.0 \pm 3.0$ & $2.41$ & $-0.3 \pm 1.9$ & $0.1 \pm 0.2$ & $\pm 0.7$ & $\pm 0.1$ & $3.0 \pm 4.1 \pm 2.7$ \\
$1.9~\rm{GeV}$ & $0.9 \pm 2.9$ & $1.70$ & $-0.2 \pm 1.3$ & $0.1 \pm 0.2$ & $\pm 0.6$ & $\pm 0.3$ & $1.4 \pm 4.0 \pm 1.9$ \\
$2.0~\rm{GeV}$ & $1.6 \pm 2.8$ & $1.10$ & $-0.2 \pm 0.9$ & $0.1 \pm 0.2$ & $\pm 0.5$ & $\pm 0.0$ & $2.2 \pm 3.8 \pm 1.3$ \\
$2.1~\rm{GeV}$ & $1.6 \pm 2.9$ & $0.65$ & $-0.1 \pm 0.5$ & $0.1 \pm 0.2$ & $\pm 0.4$ & $\pm 0.1$ & $2.2 \pm 3.9 \pm 0.9$ \\
$2.2~\rm{GeV}$ & $1.1 \pm 2.9$ & $0.38$ & $-0.1 \pm 0.3$ & $0.1 \pm 0.2$ & $\pm 0.3$ & $\pm 0.2$ & $1.4 \pm 3.9 \pm 0.6$ \\
	\hline
	\end{tabular}
	\label{table:acp:values}
\end{table*}

\newcommand{\minitab}[2][l]{\begin{tabular}{#1}#2\end{tabular}}
\begin{table*}[htbp]
  \caption{Absolute uncertainties in \acp, in percent. The systematic uncertainties are added in quadrature to yield the total.}
\centering
  \begin{tabular}{lccccccc}
    \hline
$\ecmg\rm{~thresh.}$ & Statistical & Total systematic & \adet & \abkg & MC stat. & \BBbar norm. & Wrong tag \\
 \hline
$1.70~\rm{GeV}$ & 4.26 & 3.52 & 0.30 & 3.40 & 0.76 & 0.42 & 0.02 \\
$1.80~\rm{GeV}$ & 4.13 & 2.72 & 0.30 & 2.56 & 0.68 & 0.53 & 0.05 \\
$1.90~\rm{GeV}$ & 3.96 & 1.92 & 0.30 & 1.81 & 0.58 & 0.10 & 0.02 \\
$2.00~\rm{GeV}$ & 3.84 & 1.32 & 0.30 & 1.17 & 0.48 & 0.19 & 0.04 \\
$2.10~\rm{GeV}$ & 3.91 & 0.86 & 0.30 & 0.70 & 0.39 & 0.12 & 0.04 \\
$2.20~\rm{GeV}$ & 3.89 & 0.59 & 0.30 & 0.41 & 0.30 & 0.04 & 0.03 \\
	\hline
  \end{tabular}
\label{table:uncertainties}
\end{table*}

In conclusion, we have measured the direct $CP$ asymmetry $\acp(\BSDG)$. The measurement is performed using $(772 \pm 11)\times 10^6$ \BBbar pairs for photon energy thresholds between $1.7$ and $2.2~{\GeV}$. As a nominal result we choose the $2.1~{\GeV}$ threshold since it has a low uncertainty and keeps a large fraction of signal events: $\acp(\BSDG) = (2.2 \pm 3.9 \pm 0.9) \%$, consistent with the SM prediction. This is the first Belle measurement of this asymmetry and the most precise to date.

We thank the KEKB group for excellent operation of the accelerator; the KEK cryogenics group for efficient solenoid operations; and the KEK computer group, the NII, and PNNL/EMSL for valuable computing and SINET4 network support. We acknowledge support from MEXT, JSPS and Nagoya's TLPRC (Japan); ARC (Australia); FWF (Austria); NSFC (China); MSMT (Czechia); CZF, DFG, and VS (Germany); DST (India); INFN (Italy); MOE, MSIP, NRF, GSDC of KISTI, and BK21Plus (Korea); MNiSW and NCN (Poland); MES and RFAAE (Russia); ARRS (Slovenia); IKERBASQUE and UPV/EHU (Spain); SNSF (Switzerland); NSC and MOE (Taiwan); and DOE and NSF (USA).

\end{document}